# NeuraGen-A Low-Resource Neural Network based approach for Gender Classification

Shankhanil Ghosh
School of Computer and Information Sciences
University of Hyderabad
Hyderabad, India
20mcmb04@uohyd.ac.in

Chhanda Saha
School of Computer and Information Sciences
University of Hyderabad
Hyderabad, India
20mcmb01@uohyd.ac.in

IEEE Dr. Nagamani Molakatala
School of Computer and Information Sciences
University of Hyderabad
Hyderabad, India
nagamanics@uohyd.ac.in

*Abstract*— Human voice is the source of several important information. This is in the form of features. These Features help in interpreting various features associated with the speaker and speech. The speaker dependent work researchers are targeted towards speaker identification, Speaker verification, speaker biometric, forensics using feature, and cross-modal matching via speech and face images. In such context research, it is a very difficult task to come across clean, and well annotated publicly available speech corpus as data set. Acquiring volunteers to generate such dataset is also very expensive, not to mention the enormous amount of effort and time researchers spend to gather such data. The present paper work, a Neural Network proposal as NeuraGen focused which is a low-resource ANN architecture. The proposed tool used to classify gender of the speaker from the speech recordings. We have used speech recordings collected from the ELSDSR and limited TIMIT datasets, from which we extracted 8 speech features, which were pre-processed and then fed into NeuraGen to identify the gender. NeuraGen has successfully achieved accuracy of 90.7407% and F1 score of 91.227% in train and 20-fold cross validation dataset.

## I. INTRODUCTION

Speech is a very important biometric modality to consider in today's world. The emerging of internet and multimedia has flooded the world with massive volumes of multimedia data, out which a considerable percentage is speech data. Hence, today's world has a huge demand of speech based technology. This has led to some wonderful research, such as speech activated Kisaan information system[20], voice controlled music player[19], and face-speech cross-modal matching [3][4].

However, all the referred works are very resource-hungry. Researchers need massive volumes of data to get accurate result. Arsha Nagrani et al. [9] have released a large-scale speech and face dataset for face speech cross-modal research. However, availing relevant data is still a big problem. There is a lack of well annotated data suited for the purpose of the research. Also, a lot of such well-annotated datasets are very costly. The cost of datasets range from a few hundred Euros (21,000 Indian Rupees) to even 20,000+ Euros (2,00,000+ Indian Rupees). In such cases, it becomes very difficult for non-for-profit research groups to avail such data at such high cost. Thus, in any form of research, a huge volume of time, money and resources are wasted for data collection and cleaning.

Hence, this paper work an attempt is made to develop a low-resource neural network model for gender classification from acquired speech recordings. For this work considered only publicly available and open-source datasets. The recordings in the datasets are collected from willing volunteers of both genders. The dataset has only 358 individual speech recordings, which makes it extremely low resource, as compared to other works in the related areas. The dataset was split into train and validation sets, and we have achieved 90.7407% accuracy and 91.227% F1 score in train and n-fold cross validation technique where 'n' is 20.

## II. DISCUSSION OF RELATED WORK

The Literature review for the current work is in the similar domain to get inspiration for our proposed model. Here are some of the recent works that have explored this area of speech domain.

Author Ling Feng et al. [11] proposed work is on Speech Database with the English language for Speaker Recognition (ELSDSR) database for the speaker recognition. Several datasets were available beforehand but when different feature extraction or operations were done on that the result was different for different machines. They provide the ELSDSR database pubic for any academic research in speaker recognition. The corpus is built with 22 speakers with age groups between 24 to 63. Audio features are 48 dimensional MFCC features were for their ASR experiment. To improve the recognition accuracy, applied the speaker pruning technique. With that system achieved 92.07% accuracy. Similarly J. B Millar el al.[16] have proposed Australian National Database of Spoken Language (ANDOSL) for spoken language researches in early days.

Authors Assim ara Abdulsatar et al. [2] have used MFCC and KNN to recognize the gender of audio. They have concluded that the main information of speech signal is concentrated at the low frequency part of the audio, however, the age and gender detection would give better result if they could have used SVM or ANN instead of KNN.

Sarah Ita Levitan et al. [5] work described the impact of combination of spectral and pitch feature for identifying the gender. They also worked on the cross lingual robustness of a model which is trained on English speakers to identify the gender of German speakers. Authors Saeid Safavi et al. [13] focuses on speaker, gender, and age-group recognition from children's speech. The authors compared performances of several classification methods including Gaussian Mixture Model–Universal Background Model (GMM–UBM), GMM–Support Vector Machine (GMM–SVM) and i-vector based approaches. Park et al.[14] have proposed a data augmentation method for speech recognition. The model is directly applied to the feature inputs of a NN for the task. In [10] the authors have proposed a sequence based multi-lingual low resource speech recognition model.

Z. Wang et al. [1] have proposed a feature encoding process by using Deep Neural Network (DNN) to encode individual utterances from an audio data into fixed-length vectors. The DNN encoder is jointly trained with a kernel extreme learning machine (ELM), to perform utterance-level classifier, which got improvement of 2.94% over un-weighted accuracy in age/gender classification against a strong DNN-ELM [17] approach on the emotion recognition task which is experimented on a Mandrain dataset.

Authors Zakariya Qawaqneh et al. [6] have talked about a similar cross-modal classifier that consists of two DNNs for age and gender classification from speech and facial image data. The classifier is trained on Mel-frequency cepstral coefficients (MFCC) and fundamental frequency (F0) as one feature set and shifted delta cepstral coefficients (SDC) as another feature set extracted from the utterances, and fed into the first DNN. Facial appearance and the depth information are extracted from face images as the first and second feature sets respectively for training the second DNN. The overall accuracy is 56.06% for seven speaker classes and 63.78% for Adience database. Authors Zakariya Qawaqneh et al [8] have also proposed another method for speakers age and gender classification, using the Backus-Naur Form (BNF) extractor along with DNN. The BNF extractor is used to generate transformed MFCCs features. The age/gender classification saw an improvement of 13% by using the transformed MFCC over traditional MFCC.

Authors Mingxing Duan et al. [7] have introduced a Convolution Neural Network (CNN) and Extreme Learning Machine (EML) based model to deal with age and gender classification. The CNN is used to extract the features from the input images and ELM classifies the intermediate results, which helps in preventing overfitting. Their model was tested on the MORPH-II and Adience Benchmark dataset, its accuracy and efficiency came out to be better than similar studies.

III. METHODOLOGY

A. Dataset

The entire model is trained on a very data-scarce model. The volume of data available to us was very less. The dataset that we have used for the paper is a hybrid of 2 separate datasets, both containing voice recordings by male and female subjects, of various ethnicity and age groups.

English Language Speech Database for Speaker Recognition or ELSDSR, offered by Ling Feng [11]. The audio recording for the dataset has been carried out in a room at the Technical University of Denmark. The dimensions of the room are 8.82x11.8x3.05m3. The recording is performed at the centre of the chamber, and reflectors have been used to reflect the audio. The audio files are recorded using a MARANTZ PMD670 solid state recorder. The speeches have been recorded in the *.wav format, with a sampling rate of 16kHz with a bit rate of 16 bits/sec. The dataset contains a total of 22 speakers, out of which 10 are female and 12 are male. The speakers have their ages between 24 to 63 years during the time of recording. The authors have mentioned that the demographics of the speakers, (except for gender) were not externally controlled by the researchers. The extensive details of the speakers can be found in the ELSDSR website[1].

Another batch of dataset from un-priced sample of the TIMIT Acoustic-Phonetic Continuous Speech Corpus provided by [15] is considered for empirical task. The dataset contains English language speeches from people belonging to different parts of the United States of America. It has annotations of male and female speakers. Details of the dataset can be found at the dataset website[2].

Both datasets contain speakers from mixed accents and dialects mostly from North American native English speakers and European non-native English speakers, however. However, that information has not been taken into consideration for the modeling of NeuraGen. In future improvements of this model, we might consider those features as well, which would make for an interesting study.

The proposed hybrid dataset covers the 198 audio samples from the ELSDSR [11] and 160 audio samples from the TIMIT [15] . The datasets have been coupled together to form one single speech dataset having 358 data samples. Out of this dataset we have 47% females and 52% male speakers. The dataset is pretty balanced, which would ensure that the neural network will not be biased towards any gender, and not produce erroneous prediction for any gender.

Table 1 Summary of the dataset. Each cell contains the number of speeches recording available

| S.No | Dataset name | male | female | Total |
|------|--------------|------|--------|-------|
| 1 | ELSDSR | 108 | 90 | 198 |
| 2 | TIMIT | 80 | 80 | 160 |
| 3 | Total | 188 | 170 | 358 |

B. Feature extraction and fusion methodology

After acquiring the dataset, we have performed extensive feature extraction on the dataset. We have collected a total of 8 features, namely Mel-Frequency Cepstrum Coefficient (MFCC), Root mean square (rms), Fundamental frequency, spectral-centroid, spectral bandwidth, spectral-coefficient, spectral-roll off, zero crossing rates. The features have been extracted from the raw audio samples using the Librosa 0.8.0 package using a Python 3.6.0 interpreter. We have followed a certain feature fusion mechanism on those features to generate the final feature vector which represents the audio sample. This final feature vector is then fed into NeuraGen.

The MFCC feature generates a 2-D NumPy array of dimensions (mx20) where m varies between different audio samples. We have tried several techniques of extracting features from this MFCC (mx20) array. Since each audio sample is approximately 5 seconds long, we can assume that they are quasi-static in nature, thus, the variation of the features of the MFCC features within the audio sample will not be very prominent and can be safely ignored. Hence, we have taken the mean across the columns of the MFCC matrix, and generating a final matrix of size (1x20). The feature extraction process has been visualized in Fig. 1.

---

[1] http://www2.imm.dtu.dk/~lfen/elsdsr/

[2] https://catalog.ldc.upenn.edu/LDC93S1/

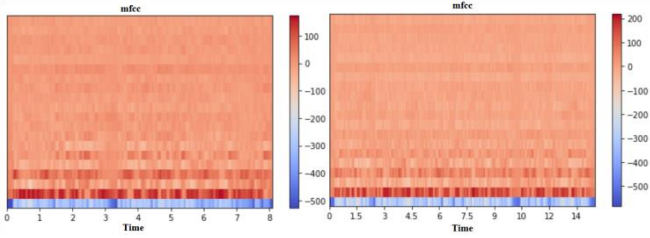

Fig. 1 MFCC visualization of audio samples by male (right) and female (left) speakers, from TIMIT dataset.

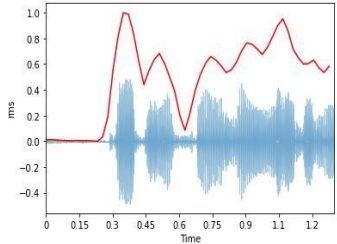

Fig. 2a RMS visualization

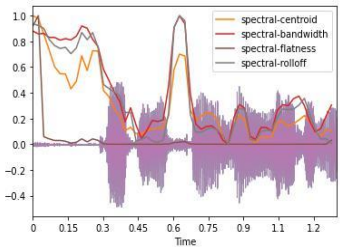

Fig. 2b Spectral features visualization

Fig. 2 RMS and Spectral features visualization of an audio sample by a male speaker, ID = MCBR from TIMIT dataset.

The remaining 7 features, RMS (Fig. 2a), fundamental frequency, Spectral-features (Fig. 2a) and ZCR are 1-D feature vectors of dimension (1xm), represented against time. The value of m varies due to the length of the audio sample. In a similar approach mentioned above, we have taken the mean of the entire feature vectors for each of those features and converted them into (1x1) features.

After performing the necessary feature extraction, we concatenated the features along the length of the vector, to generate one single feature vector of dimension (1x27), In the feature vector, the first (1x20) sub-vector represented the mean-MFCCs, and the rest 7x (1x1) features represented the 7 mean of feature vectors as mentioned above, in that same order. The total dataset, when represented as a matrix, was of dimensions 358x27. The feature extraction and fusion mechanism has been visualized in Fig. 3.

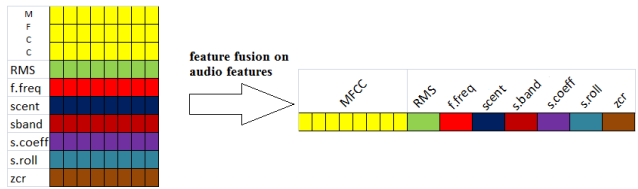

Fig. 3 Feature extraction and fusion mechanism

### C. NeuraGen architecture

The NeuraGen is a low resource neural network, which takes in the audio feature vector, which we obtained in the previous section, and outputs the probability of the audio belonging to either Male or female. Since the gender information is binary in nature, we have represented Female gender as 1, and Male gender as 0. The output layer of NeuraGen (which has a softmax activation function) generates a real probability value in the range [0, 1], which rounds off to either 0 or 1, indicating male or female. The NeuraGen has the following architecture:

1) Input layer of shape (1x27), which is the audio feature vector
2) Hidden layer 1 of shape (1x25), with ReLU activation function in each unit.
3) Hidden layer 2 of shape (1x10), with ReLU activation function in each unit.
4) Hidden layer 3 of shape (1x5), with ReLU activation function in each unit
5) Output layer having 1 unit, with Sigmoid activation function

Since the NeuraGen is a low resource neural network, hence it suffers from over fitting. To deal with the problem of over fitting, we have added additional Dropout regularization layers for the first two hidden layers, with rate=0.01. These Dropout layers will disable certain perceptrons from the hidden layers as needed to prevent overfitting. The model will be publicly distributed by the authors after some further tuning of the hyper-parameters. The entire architecture of NeuraGen is given in Fig. 5, which is an extension of the feature fusion and extraction as shown in Fig. 3.

## IV. EXPERIMENT SETUP AND RESULTS

### A. Experiment setup

As mentioned before, the hybrid dataset that we have taken for this experiment contains 358 data samples. We have stored the input dataset in CSV file format, in a way that would make it easy and legible for NeuraGen to read and process. The dataframe format is given in Fig. 4. The first cell is the serial number of the data, ranging from 0 to 357. The next 27 columns include the (1x27) feature vector for the $i^{th}$ audio data, and the last column contains the information about the audio belonging to a male or a female voice. The last cell of the dataframe (which contains the gender information) is a categorical data of string type, consisting of 4 possible values: Male, M for male speaker audio and Female, F for female-speaker audio.

We have used Pandas 1.2.4 to work with the dataframe format of the dataset.

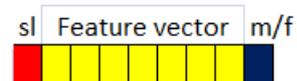

Fig. 4 Dataframe format for storing the data

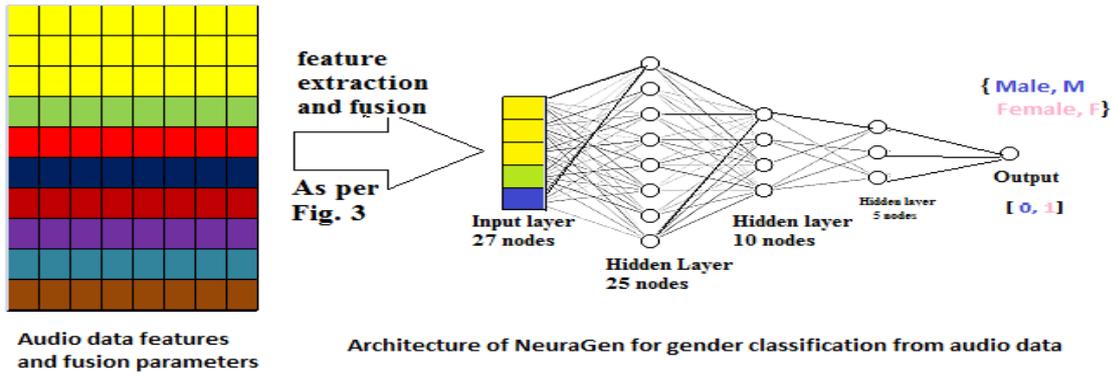

Fig. 5 NeuraGen architecture

The entire feature vectors of all the audio data is represented as a matrix of 358x27. However this feature matrix is prepared by concatenation of various features, and hence the range of the data in the matrix varies vastly. For example, the mean MFCC value vary from approximately [-400, +400], and they heavily fluctuate within the feature vector itself, (the range is not exact, only an approximate), while the RMS value can range from 0 to 1. The fundamental frequency can be of range 2000-3000Hz (humans can percept audio from frequency of 20Hz to 20 kHz). It can be clearly seen that there is a very large disparity between the ranges of the values within the feature, which will cause the neural network to work poorly. Hence, we have normalized the feature values, to generate a normalized feature matrix.

Also, as we have mentioned previously, the male and female values are represented by 4 categorical string values, which we have to convert into binary valued categorical variable. Thus, we have replaced "F" and "Female" into 1, and "M" and "Male" into 0.

We have segregated the dataset into 2 parts, the training subset and the validation subset. We have considered 2 different experiments, one with 10 fold cross validation and another with 20 fold cross validation. In each case, the train and validation datasets have been randomly distributed.

### B. Model training and experiment results

After preprocessing the dataset, an attempt made to train the neural network using the defined dataset. The training is carried out by the neural network over various epochs duration, and then found, that the neural network reaches optimal loss and accuracy value at approximately 500 epochs. Hence, recommended that NeuraGen and any of its variation be trained over 500 epochs. Epoch over 500 will lead to over fitting, and lesser epoch will lead to under fitting. Here 4 different performance metrics are used to understand the performance of proposed model, namely, *Accuracy, Loss*, *Precision* and *Recall*.

The Scores are collected after multiple iterations of training and cross-validation, to make sure that the metrics that received are not achieved by chance. Also an attempt is made to test our model with different sizes of training and cross validation sub-dataset. For this 3 such sizes are used, namely 5-fold, 10-fold and 20-fold. We have plotted the learning curves for the training and the cross-validation sub-datasets, for all 4 types of metrics. Table 2 gives a tabular view of the metrics values for 3 types of training and cross-validation sub-datasets. We have highlighted the best performance metrics for each case in bold. Fig. 6 shows the accuracy and loss training and Fig. 7 shows the precision and recall for 3 types of sub-dataset.

Table 2 NeuraGen performance metrics over train and cross validation datasets

| S.No | Dataset | Accuracy | Loss | Precision | Recall | F1-score |
|---|---|---|---|---|---|---|
| 1 | 20-fold CV | **0.9074** | **0.1721** | **0.9629** | 0.8666 | **0.9122** |
| 2 | 10-fold CV | **0.9074** | 0.1960 | 0.9166 | **0.8799** | 0.8979 |
| 3 | 5-fold CV | 0.8888 | 0.2695 | 0.8888 | 0.9411 | **0.9142** |

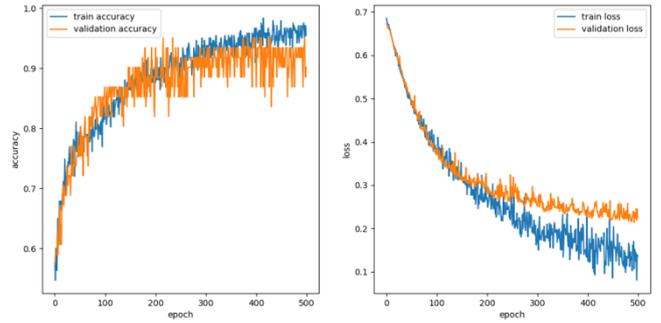

Fig. 6a For 20-fold cross validation

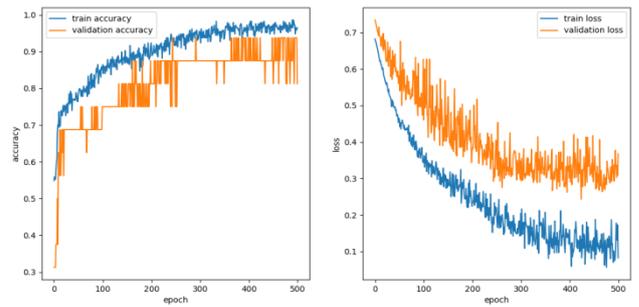

Fig. 6b For 10-fold cross validation

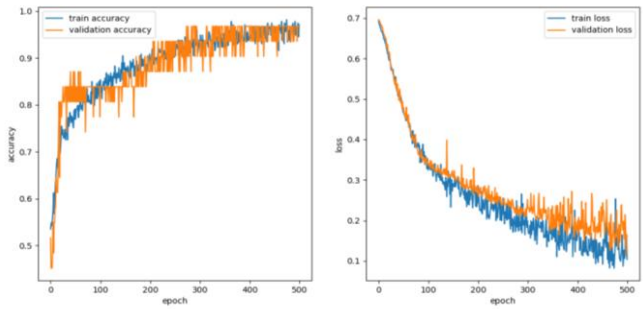

Fig. 6c For 5-fold cross validation

Fig. 6 Accuracy and loss training curves for training and 20-fold, 10-fold and 5-fold cross validation datasets

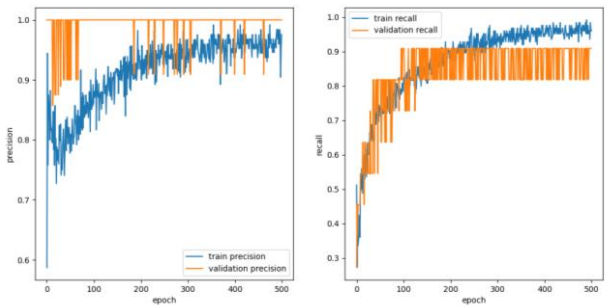

Fig. 7a for 20-fold cross validation

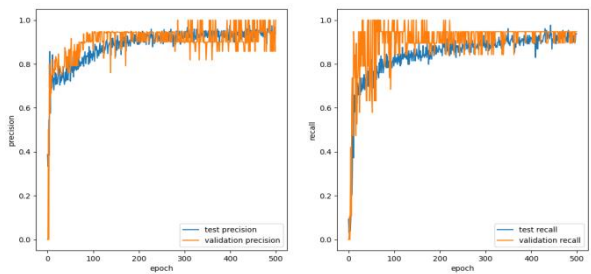

Fig. 7b for 10-fold cross validation

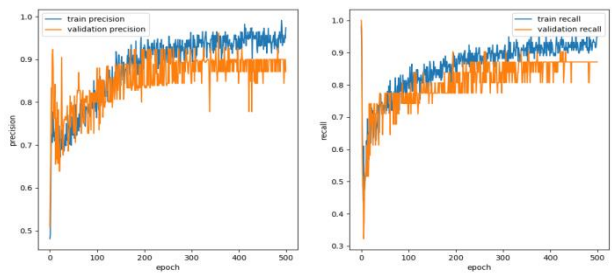

Fig. 7c For 5-fold cross validation

Fig. 7 Precision and recall training curves for training and 20-fold, 10-fold and 5-fold cross validation datasets

### C. Analysis of experiment results

According to Table 2 the precision of NeuraGen for 20-fold cross validation is above 96%, which is a significant edge over the other 2 types of dataset. However the recall is the lowest at 86.66%. Table 2 also suggests that the 20-fold cross validation provides a much better performance in almost all respects, in comparison to with 10-fold or 5-fold cross validation sub-sets. Low resource models need some a good volume of cross-validation dataset to make sure the model is not over fitting, which can be a big problem in case of low-resource models.

We have also compared our model with other models that have been proposed previously. For our study, we have taken 3 models to compare, namely, DNN-ELM model proposed by Z. Wang et. Al. [1], BNF-DNN model, proposed by Zakariya Qawaqneh et. al. [8] and Fine tuned DNN proposed by Zakariya Qawaqneh et. at. [6]. All these models are deep learning models tested over very high volume of dataset. The purpose of this study is to show how our low-resource NeuraGen performs in comparison to deep learning approaches with high-resource dataset. We have only taken the accuracy of the models into consideration for this comparison.

Table 3 Comparison of NeuraGen with other Gender classification models

| S. No | Approach | Dataset | Dataset volume (hours) | Accuracy% | Target classification |
|---|---|---|---|---|---|
| 1. | DNN-ELM[1] | IEMOCAP [12] | 12 | 92.72 | Age+Gender |
| 2. | BNF-DNN[8] | AGender [18] | 47 | 58.98 | Age+Gender |
| 3. | Fine tuned DNN [6] | AGender [18] | 47 | 56.06 | Age+Gender |
| 4. | NeuraGen | TIMIT [15] + ELSDSR [11] | ≈ 0.5 | 90.74 | Gender |

The comparison has been shown in Table 3. As can be seen from Table 3, NeuraGen uses the smallest volume of data, with approximately 30 minutes of data, whereas the other models, which employ deep learning models, have 12-47 hours of data. NeuraGen has performed significantly good in comparison to [6], [8], and has performed fairly good, when compared with [1]. From this comparison, we can clearly see that NeuraGen has been able to perform gender classification with a fairly good accuracy, even with extremely low volume of dataset.

### V. CONCLUSION

For this work, a proposal is made for NeuraGen which is a low-resource ANN model for gender classification. Here, an attempt is made to develop speech models with very limited volume of data available. For this purpose acquired male and female speech recordings from 2 datasets where willing volunteers recorded their voice by reading from a sample text. We have performed feature extraction on 358 audio samples, from which extracted 8 audio features, and performed feature fusion to convert each audio sample into 1x27 size feature vectors, which were then fed into a fully connected neural network. NeuraGen has achieved accuracy of 90.7404% and F1-score of 91.227% on train and 20-fold cross-validation dataset. We have also compared our models with other gender-classification models and have shown that our model works fairly well with only approximately 30 minutes of audio data. This model is a proof that it is possible to achieve a relatively high accuracy in a low-resource biometric-classification model.

### VI. FUTURE SCOPE OF WORK

The purpose of NeuraGen is to extract relevant gender information from an audio-sample, which will then be used as a feature in subsequent machine learning tasks. This model is by no means the best model when it comes to low-resource speech models. There is a big area left for improvement in terms of performance metrics. Another big limitation to this model is that, we have not tested how this model works for noisy, or out-of-lab audio samples. Also, this model has been trained and tested over a very specific demographic, and we do not have relevant data to test how the change in demographics affects the performance of the model. In our future improvements of NeuraGen, we shall look into these aspects and improve the model's performance. One major focus of this research is to develop models which can be used in out-of-lab environments, and hence susceptibility to noisy audio is a must. We would also test different types of approaches to biometric classification (including gender/age detection), using transfer learning, for low-resource models.

ACKNOWLEDGMENT

The authors would like to express their gratitude towards the School of Computer and Information Sciences, University of Hyderabad for supporting the research, even in these trying times of a global pandemic. The infrastructure and technical support provided by them has made this research easier. We would also like to thank Ms. Ling Feng of Neurobiology Research Unit, Rigs Hospitalet, Copenhagen University Hospital for providing access to the ELSDSR dataset, which has made this research possible.